# PARTITION FUNCTION FOR TWO-DIMENSIONAL NEAREST NEIGHBOUR ISING MODEL IN A NON-ZERO MAGNETIC FIELD FOR A SQUARE LATTICE OF 16 SITES

Nandhini.G and Sangaranarayanan.M.V *

Department of Chemistry

Indian Institute of Technology -Madras, Chennai – 600036

India

Email : sangara@iitm.ac.in

Fax : +9144 22570545

**Abstract**

An explicit expression for the partition function of two-dimensional nearest neighbour Ising models in the presence of a magnetic field is derived by a systematic enumeration of all the spin configurations pertaining to a square lattice of sixteen sites. The critical temperature is shown to be in excellent agreement with the values of Yang and Lee while the corresponding 'dimensionless' magnetic field is obtained as 0.004.

---------------------------------------------------------------------------------------------------------

The analysis of Ising models constitutes a central theme in statistical mechanics and its applicability ranges from the study of critical behaviour of ferromagnets [1] to temperature-dependence of base pair sequences in DNA [2]. While the one-dimensional nearest neighbour Ising model did not exhibit phase transitions, the exact solution of the two-dimensional Ising model for zero magnetic field reported by Onsager[3] yielded an explicit expression for critical temperatures. Lee and Yang [4] provided a complete solution of the two-dimensional Ising model for vanishing magnetic fields, validating the exact solution of Onsager. However, subsequent attempts of analyzing the two-dimensional Ising models when the magnetic field is non-zero has remained elusive till now, although methodologies employing Bragg-Williams approximation [5] , Bethe *ansatz* [6], series expansions [7] , renormalization group [8]*,* scaling hypothesis [9], etc. have  been investigated. In view of its equivalence with binary alloys and lattice gas description of fluids [10], the results derived in the context of Ising models become applicable *mutatis mutandis* to various topics in solid state and condensed matter physics. In contrast to the one-dimensional case, the analysis of two-dimensional Ising models employing the transfer matrix method is considered almost impossible since the total number of spin configurations for a square lattice of N sites is $2^N$ and even for N=16, this number equals 65,536. The two-dimensional Ising model [11] finds applicability in the study of order-disorder transitions [12], electrochemical interfaces [13], phase separation in self-assembled monolayer films [14], protein folding [15], free energies of surface steps [16] etc. As pointed out elsewhere [17] "In terms of the number of papers published, Ising model ranks as probably the most celebrated model in mathematical physics.."

In this Report, we derive an explicit relation for the partition function of two-dimensional nearest neighbor Ising models for a square lattice of N sites in the presence of magnetic field (H). This equation for the limiting case of H = 0 yields partition functions in agreement with those arising from Onsager's exact solution. A new method of estimating the critical temperature from the partition function is also suggested and an expression for the spontaneous magnetization is proposed.

**Methodology**

Consider the two-dimensional nearest neighbour Ising model on a square lattice with the Hamiltonian given by

$$H_T = -J\sum_{<ij>}(\sigma_{i,j}\sigma_{i,j+1} + \sigma_{i,j}\sigma_{i+1,j}) - H\sum_{i,j}\sigma_{i,j} \qquad (1)$$

where J is the nearest neighbour interaction energy, H being the external magnetic field. i and j denote the row and column index respectively. The periodic boundary conditions are assumed. The corresponding canonical partition function is defined as

$$Q = \mathrm{Tr}\{\exp(-H_T/kT)\} \qquad (2)$$

where k denotes the Boltzmann constant, T being the absolute temperature. For brevity, we represent as $e^{J/kT}$ as x and $e^{H/kT}$ as y; J and H are assumed to be positive quantities.

Our *tour de force* consists in deducing the partition function for a square lattice of 16 sites by a systematic enumeration of all the 65,536 configurations. Since the partition function can also be formulated as $Q = \sum_{i=1}^{2^N} e^{-E_i/kT}$, the sum of all the 65536 terms involving the energies yields Q. This strategy avoids (i) the construction of transfer matrices and (ii) diagonalization of matrices followed by a search for dominant eigenvalues.

In view of this enumeration, the analysis gets simplified as a summation of all the 65536 'energy' terms. The sum $\sum_{i=1}^{65536} e^{-E_i/kT}$ yields the canonical partition function.

**Partition function for H=0**

Before deducing the partition function for H ≠ 0, we derive the result for H = 0, since the exact solution of Onsager is available in this case. The complete algebraic expression for the partition function is as follows:

$Q_{H=0}$= ($8x^{18}$ +$144x^{-6}$+860+ $12x^{-12}$+ $40x^{-10}$+ $64x^4$+ $32x^{-12}$+ $8x^{-18}$ + $8x^{-16}$+ $808x^8$+ $1184x^2$+$704x^{-4}$+$56x^{18}$+$112x^{16}$+ $24x^{18}$+ $144x^{12}$+$208x^{14}$+$8x^{20}$+$24x^{18}$+ $104x^{14}$+ $8x^8$+$8x^{20}$+$40x^{-10}$+$2x^{24}$+$64x^{-4}$+$2y^8$+$8x^{20}$+280+$96x^{14}$+$112x^{14}$+$90x^{16}$+$920x^{-4}$+ $248x^{12}$+ $38x^{16}$+$x^{32}$+$16x^{16}$+$40x^2$+$1564x^4$+$8x^{24}$ +$8x^{22}$+1592+$8x^2$+$8x^{24}$+ $32x^{20}$+ $8x^{22}$+ $2x^{24}$+$8x^{24}$+$368x^{-6}$+$2x^{24}$+$16x^{16}$+$96x^{10}$+ $176x^4$+$40x^2$+$8x^{-6}$+$56x^{18}$+$8x^{20}$ + $120x^{14}$+ $296x^{12}$+$90x^{16}$+$24x^{18}$+$152x^{14}$+$8x^{20}$+$40x^{16}$+$168x^{-8}$+$16x^{20}$+$888x^6$+$208x^{14}$+ $96x^{-10}$+$2x^{-24}$+$176x^4$+$8x^{22}$+ $40x^{16}$+$352x^{12}$+$8x^{18}$+$400x^{10}$+ $352x^{12}$ +$808x^8$+ $112x^{14}$+ $8x^{20}$+$536x^{10}$+ $1264x^6$+$536x^{10}$+$36x^{16}$+$88x^{-10}$+$8x^{-16}$+$8x^{-14}$+$112x^{-2}$+$1936x^2$+$900x^8$+ $656x^{10}$+ 860+ 1592+ $2152x^4$+$1184x^2$+$1056x^8$+$464x^2$+$1512x^6$+$72x^{-12}$+$350x^8$+ $888x^6$+$656x^{10}$+$224x^6$+$1264x^6$+ $272x^{12}$+$632x^4$+$900x^8$+$104x^{14}$+$1564x^4$+ $584x^{10}$+$96x^{10}$+$16x^{20}$+$56x^{18}$+$120x^{14}$+$36x^{-8}$+$8x^{22}$+$x^{32}$+$144x^{-6}$+$224x^6$+280+ $152x^{14}$+$36x^{-8}$+1770+$16x^{16}$+2+$32x^{-12}$+$16x^{-14}$+ $464x^2$+$400x^{10}$+$144x^{12}$+ $32x^{20}$+$1472x^{-2}$+$1864x^4$+$1512x^6$+$300x^{12}$+$250x^{-8}$+$624x^{-2}$+$112x^{-2}$+ $368x^{-6}$+$1104x^{-2}$+$704x^{-4}y^2$+$8x^{-6}y^6$+$8x^{20}$+$8x^{20}$+$8x^8$+$12x^{-12}$+$624x^{-2}$+$260x^{-4}$+$632x^4$+$168x^{-8}$+ $1104x^{-2}$+$448x^{-6}$+ $2096x^2$+$8x^{24}$+$88x^{-10}$+$8x^{-14}$+ $56x^{18}$+$296x^{12}$+$16x^{16}$+$350x^8$+ $112x^{16}$+ $584x^{10}$+$36x^{16}$+ $300x^{12}$+$8x^{18}$+$1056x^8$+$24x^{18}$+$608x^{10}$+$994x^8$+$1568x^6$+$8x^{-18}$+ $260x^{-4}$+$1936x^2$+$1864x^4$+ $272x^{12}$+$8x^{26}$+$22x^{-16}$+$8x^{18}$) (3)

The partition functions predicted by the above equation are compared with those arising from Onsager [3] in Fig 1 for values of J/ kT ranging from 0 to 2.0. A very good agreement can be noticed indicating the validity of the above equation for H = 0.

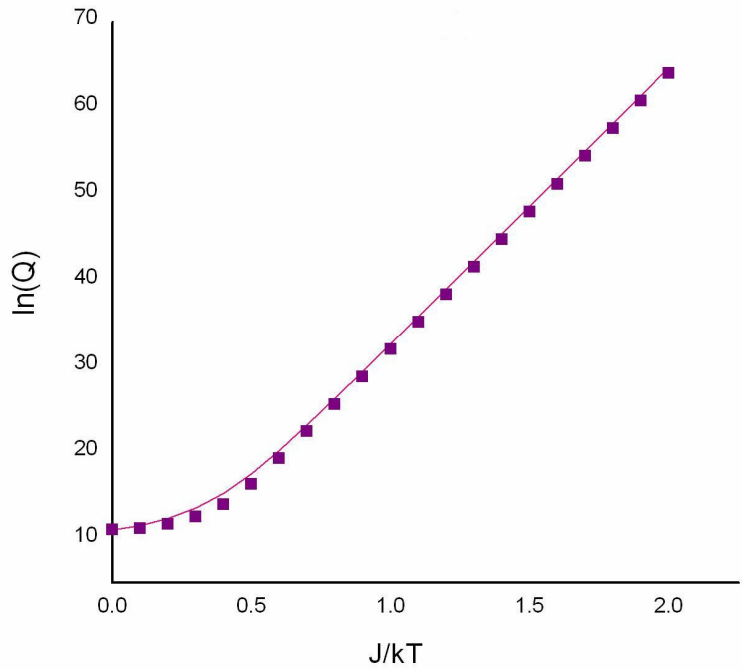

Figure 1 - A comparison of the partition function values estimated from equation (3) with that of Onsager's analysis [3]. Line denotes the estimates of equation (3) while the symbols denote the values obtained from Onsager's exact solution.

**Partition function in the presence of a magnetic field**

Since the above methodology arising from the enumeration of 65,536 spin configurations has yielded partition functions in agreement with Onsager's exact solution for H=0, the spin arrangements are anticipated to be correct for $H \neq 0$ or equivalently $y \neq 1$. The summation of the energy terms yields the partition function as

$Q_{H\neq 0}=(8x^{18}y^{-4}+144x^{-6}y^4+860y^{-4}+12x^{-12}y^{-4}+40x^{-10}y^{-4}+64x^4y^{-6}+32x^{-12}y^{-2}+8x^{18}y^{-2}+8x^{-16}y^{-2}+808x^8y^{-6}+1184x^2y^{-4}+704x^{-4}y^{-2}+56x^{18}y^{12}+112x^{16}y^{10}+24x^{18}y^8+144x^{12}y^{10}+208x^{14}y^8+8x^{20}y^6+24x^{18}y^{-8}+104x^{14}y^{-2}+8x^8y^{10}+8x^{20}y^{-2}+40x^{-10}y^4+2x^{24}+64x^{-4}y^6+2y^8+8x^{20}y^{-8}+280y^6+96x^{14}+112x^{14}y^{-4}+90x^{16}y^{-8}+920x^{-4}+248x^{12}+38x^{16}+x^{32}y^{16}+16x^{16}y^2+40x^2y^8+1564x^4y^{-4}+8x^{24}y^{-14}+8x^{22}y^{12}+1592y^{-2}+8x^26y^{14}+8x^{24}y^{12}+32x^{20}y^{12}+8x^{22}y^{10}+2x^{24}y^8+8x^{24}y^{14}+368x^{-6}y^{-2}+2x^{24}y^{-8}+16x^{16}y^{12}+96x^{10}y^{10}+176x^4y^8+40x^2y^{-8}+8x^{-6}y^{-6}+56x^{18}y^{10}+8x^{20}y^8+120x^{14}y^{10}+296x^{12}y^8+90x^{16}y^8+24x^{18}y^6+152x^{14}y^6+8x^{20}y^4+40x^{16}y^{-6}+168x^{-8}y^2+16x^{20}y^{10}+888x^6y^{-6}+208x^{14}y^{-8}+96x^{-10}+2x^{-24}+176x^4y^{-8}+8x^{22}y^{-12}+40x^{16}y^6+352x^{12}y^6+8x^{18}y^4+400x^{10}y^8+352x^{12}y^{-6}+808x^8y^6+112x^{14}y^4+8x^{20}y^2+536x^{10}y^{-6}+1264x^6y^{-4}+536x^{10}y^6+36x^{16}y^4+88x^{-10}y^2+8x^{-16}y^2+8x^{-14}y^{-2}+112x^{-2}y^{-6}+1936x^2y^{-2}+900x^8y^{-4}+656x^{10}y^{-4}+860y^4+1592y^2+2152x^4+1184x^2y^4+1056x^8y^2+464x^2y^6+1512x^6y^2+72x^{-12}+350x^8y^8+888x^6y^6+656x^{10}y^4+224x^6y^8+1264x^6y^4+272x^{12}y^2+632x^4y^6+900x^8y^4+104x^{14}y^2+1564x^4y^4+584x^{10}y^2+96x^{10}y^{-10}+16x^{20}y^{-10}+56x^{18}y^{-10}+120x^{14}y^{-10}+36x^{-8}y^{-4}+8x^{22}y^{-10}+x^{32}y^{-16}+144x^{-6}y^{-4}+224x^6y^{-8}+280y^{-6}+152x^{14}y^{-6}+36x^{-8}y^4+1770+16x^{16}y^{-12}+2y^{-8}+32x^{-12}y^2+16x^{-14}+464x^2y^{-6}+400x^{10}y^{-8}+144x^{12}y^{-10}+32x^{20}y^{-12}+1472x^{-2}+1864x^4y^{-2}+1512x^6y^{-2}+300x^{12}y^{-4}+250x^{-8}+624x^{-2}y^4+112x^{-2}y^6+368x^{-6}y^2+1104x^{-2}y^2+704x^{-4}y^2+8x^{-6}y^6+8x^{20}y^{-4}+8x^{20}y^{-6}+8x^8y^{-10}+12x^{-12}y^4+624x^{-2}y^{-4}+260x^{-4}y^{-4}+632x^4y^{-6}+168x^{-8}y^{-2}+1104x^{-2}y^{-2}+448x^{-6}+2096x^2+8x^{24}y^{-12}+88x^{-10}y^{-2}+8x^{-14}y^2+56x^{18}y^{-12}+296x^{12}y^{-8}+16x^{16}y^{-2}+350x^8y^{-8}+112x^{16}y^{-10}+584x^{10}y^{-2}+36x^{16}y^{-4}+300x^{12}y^4+8x^{18}y^2+1056x^8y^{-2}+24x^{18}y^{-6}+608x^{10}+994x^8+1568x^6+8x^{-18}y^2+260x^{-4}y^4+1936x^2y^2+1864x^4y^2+272x^{12}y^{-2}+8x^{26}y^{-14}+22x^{-16}+8x^{18}y^{-2})$                  (4)

Although the above equation when H= 0 has shown excellent agreement with the partition functions predicted by Onsager's analysis(Fig 1), there exists no direct method for verifying the above equation since the exact partition function for two-dimensional nearest neighbour Ising models in the case of non-vanishing magnetic fields has hitherto not been reported. While the series expansions provide the partition functions, employing them would *inter alia* require defining the 'temperature regimes' (high and low) and hence the comparison with series expansions is not attempted here. In Fig 2, we report the partition function estimates derived from eqn (4) for different interaction energies and magnetic fields.

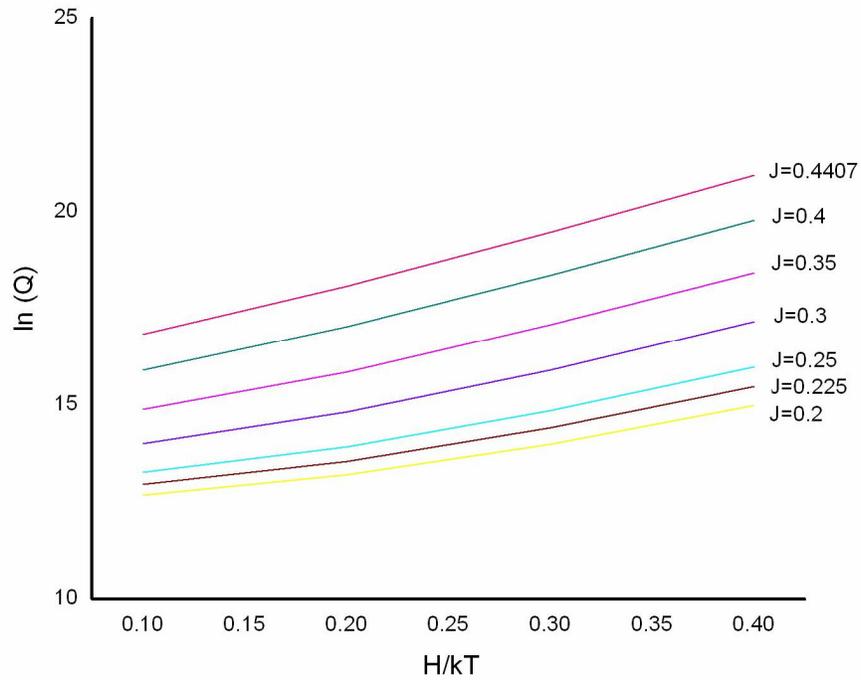

Figure 2 - The dependence of the partition function on the magnetic field for different nearest neighbour interaction energies estimated using eqn (4).

**Estimation of critical temperature**

### (A) From Spontaneous Magnetization

The precise definition of the spontaneous magnetization for two-dimensional Ising model is subtle. A commonly employed method of estimating the same is [10]

$$M_{H=0} = \lim_{N \to \infty} \frac{1}{N} \left( \frac{d \ln(Q)}{dH} \right)_{H \to 0^+} \tag{5}$$

where (i) the order of evaluating the limits is crucial and (ii) the thermodynamic limit $_{N \to \infty}$ needs to be invoked. Since the present analysis has assumed a finite system with a small number of sites, the above definition will not be directly applicable. However, it is well known that at $J/kT = 0$, the spontaneous magnetization is unity. Further M should be directly related to $\left( \frac{d \ln Q}{dH} \right)$ and hence a systematic search for the parametric dependence of M on $\left( \frac{d \ln Q}{dH} \right)$ led to

$M_{H=0.004} =$

$1.128 - 2 \times ((538.6980x^{10} + 567.3559x^8 + 15.6212x^{22} + 580.6595x^6 + 170.7921x^{16} + 532.01x^4 + 415.3001x^{12} + 57.3625x^{20} + 263.9846x^{14} + 0.2560x^{-18} + 129.8315x^{18} + 367.6395x^2 + 22.7943x^{24} + 2.0494x^{32} + 7.9362x^{-10} + 12.5506x^{26} + 0.2560x^{-16} + 74.2434x^{-4} + 0.2560x^{-14} + 242.7012 + 147.4629x^{-2} + 32.5131x^{-6} + 2.5601x^{-12} + 9.9843x^{-8})/ (5153.1x^{10} + 7239.1x^8 + 32.0312x^{22} + 935.2x^6 + 658.3415x^{16} + 10625x^4 + 2976.8x^{12} + 160.1147x^{20} + 1488.5x^{14} + 16.0005x^{-18} + 352.2596x^{18} + 9344.7x^2 + 38.0456x^{24} + 2.0041x^{32} + 352.0159x^{-10} + 16.0251x^{26} + 38.0005x^{-16} + 2976.1x^{-4} + 32.0005x^{-14} + 7238.5 + 5152.3x^{-2} + 1488.1x^{-6} + 160.0051x^{-12} + 2x^{-24} + 658.020x^{-8}))$ (6)

where x = $e^{J/kT}$. Interestingly, solving the above equation with the help of MATLAB confirms that there exists *only one real root* of equation (6) at J/kT = 0.4407.

The latter value is in excellent agreement with the critical temperature of two-dimensional nearest neighbour square lattice Ising model derived by Yang and Lee [4] in the thermodynamic limit. The behaviour of M as a function of $T/T_c$ is shown in Fig 3 wherein the magnetization vanishes at the critical temperature. We note that a complete analysis of magnetization using thermodynamic limit of N→ ∞ predicts the phase transition at H = 0 and J/kTc = 0.4407. In contrast, this investigation yields H/kT as 0.004 and J/kTc = 0.4418. The fact that our analysis has yielded a value of 4×10$^{-3}$ for 'critical magnetic field' in stead of the exact value of zero while the value of J/kTc is predicted to be 0.5×log (2.4195) i.e.0.4418 in stead of $\left[1-\sinh^{-4}(2J)\right]^{1/8}$ or (0.4407) arising from strict thermodynamic analysis, indicates the extent to which the finite lattice sizes can be subjected to in the analysis of critical phenomena.

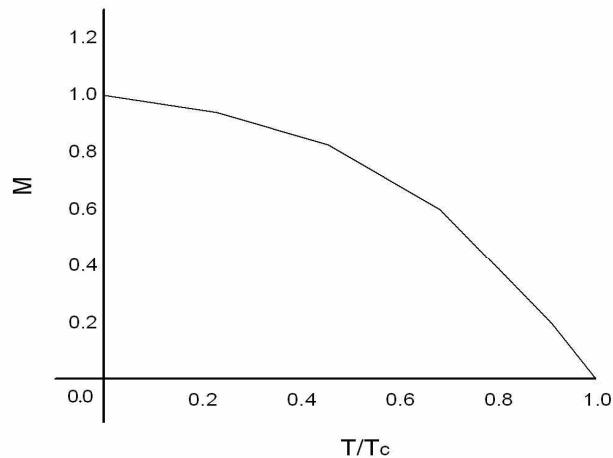

Figure 3 - The spontaneous magnetization evaluated using eqn (6) at different values of J/kT for H= 4×10$^{-3}$.

**(B) From Partition function**

During the search for critical temperatures arising from the magnetization expression, an entirely novel method of obtaining the same from the partition function at H = 0 itself was deduced by us, on account of the availability of an algebraic expression (eqn 3). Using eqn (3), it was observed that the quantity

$$\frac{1}{16}\ln(Q) - 1 = \frac{1}{16} \times \log(8x^{18} + 144x^{-6} + 860 + 12x^{-12} + 40x^{-10} + 64x^4 + 32x^{-12} + 8x^{-18} +$$
$$8x^{-16} + 808x^8 + 1184x^2 + 704x^{-4} + 56x^{18} + 112x^{16} + 24x^{18} + 144x^{12} + 208x^{14} + 8x^{20} + 24x^{18} +$$
$$104x^{14} + 8x^8 + 8x^{20} + 40x^{-10} + 2x^{24} + 64x^{-4} + 2y^8 + 8x^{20} + 280 + 96x^{14} + 112x^{14} + 90x^{16} +$$
$$920x^{-4} + 248x^{12} + 38x^{16} + x^{32} + 16x^{16} + 40x^2 + 1564x^4 + 8x^{24} + 8x^{22} + 1592 + 8x^2 +$$
$$8x^{24} + 32x^{20} + 8x^{22} + 2x^{24} + 8x^{24} + 368x^{-6} + 2x^{24} + 16x^{16} + 96x^{10} + 176x^4 + 40x^2 + 8x^{-6} +$$
$$56x^{18} + 8x^{20} + 120x^{14} + 296x^{12} + 90x^{16} + 24x^{18} + 152x^{14} + 8x^{20} + 40x^{16} + 168x^{-8} + 16x^{20} +$$
$$888x^6 + 208x^{14} + 96x^{-10} + 2x^{-24} + 176x^4 + 8x^{22} + 40x^{16} + 352x^{12} + 8x^{18} + 400x^{10} + 352x^{12} +$$
$$808x^8 + 112x^{14} + 8x^{20} + 536x^{10} + 1264x^6 + 536x^{10} + 36x^{16} + 88x^{-10} + 8x^{-16} + 8x^{-14} + 112x^{-2} +$$
$$1936x^2 + 900x^8 + 656x^{10} + 860 + 1592 + 2152x^4 + 1184x^2 + 1056x^8 + 464x^2 + 1512x^6 +$$
$$72x^{-12} + 350x^8 + 888x^6 + 656x^{10} + 224x^6 + 1264x^6 + 272x^{12} + 632x^4 + 900x^8 + 104x^{14} +$$
$$1564x^4 + 584x^{10} + 96x^{10} + 16x^{20} + 56x^{18} + 120x^{14} + 36x^{-8} + 8x^{22} + x^{32} + 144x^{-6} +$$
$$224x^6 + 280 + 152x^{14} + 36x^{-8} + 1770 + 16x^{16} + 2 + 32x^{-12} + 16x^{-14} + 464x^2 + 400x^{10} +$$
$$144x^{12} + 32x^{20} + 1472x^{-2} + 1864x^4 + 1512x^6 + 300x^{12} + 250x^{-8} + 624x^{-2} + 112x^{-2} + 368x^{-6} +$$
$$1104x^{-2} + 704x^{-4}y^2 + 8x^{-6}y^6 + 8x^{20} + 8x^{20} + 8x^8 + 12x^{-12} + 624x^{-2} + 260x^{-4} + 632x^4 + 168x^{-8} +$$
$$1104x^{-2} + 448x^{-6} + 2096x^2 + 8x^{24} + 88x^{-10} + 8x^{-14} + 56x^{18} + 296x^{12} + 16x^{16} + 350x^8 +$$
$$112x^{16} + 584x^{10} + 36x^{16} + 300x^{12} + 8x^{18} + 1056x^8 + 24x^{18} + 608x^{10} + 994x^8 + 1568x^6 +$$
$$8x^{-18} + 260x^{-4} + 1936x^2 + 1864x^4 + 272x^{12} + 8x^{26} + 22x^{-16} + 8x^{18}) - 1 \qquad (7)$$

is zero when $J/kT = 0.4300$. This value is in ca 2% error in comparison with the value of 0.4407 expected for the critical temperature when the spontaneous magnetization becomes zero. This was found to be true using Onsager's exact solution for any square lattice of N sites wherein $[(1/N) \ln Q] - 1 = 0$ at $T_c$. This feature indicates that it may be possible to deduce the critical temperature solely from the partition function at H=0 for two-dimensional nearest neighbour Ising models.

In Summary, an explicit expression for computing the partition function for the two-dimensional nearest neighbour Ising model in the case of a non-vanishing magnetic field is derived by counting all the spin configurations for a square lattice of sixteen sites. When the magnetic field is zero, the partition functions predicted by Onsager's exact solution are obtained. The critical parameters are deduced as $H/kT = 0.004$ and $J/kT = 0.4418$ for this system. A new method of estimating the critical temperature from the partition function is also proposed.

**REFERENCES**

1. Chamberlin, R.V. Mean-field model for the critical behaviour of ferromagnets. *Nature* **408,** 337-339 (2000).

2. Peryard, M. Melting the double helix, *Nature Physics* **2**, 13-14 (2006).

3. Onsager,L. Crystal Statistics. I. A Two-Dimensional Model with an Order-Disorder Transition. *Phys.Rev.* **65**,117-149 (1944).

4. Lee,T.D. & Yang, C.N. Statistical theory of equations and phase transitions. II. Lattice gas and Ising Model. *Phys. Rev.* **87 ,** 410 (1952).

5. Bragg, W.L. & Williams,E.J. The effect of thermal agitation on atomic arrangement in alloys. *Proc. Roy. Soc (London)* **A145**, 699-730 (1934).

6. Bethe, H.A. Statistical theory of superlattices. *Proc. Roy. Soc (London)* **A150**, 552-575 (1935).

7. Domb, C. In Domb, C. & Green, M. S., eds. *Phase Transitions and Critical Phenomena*. **3,** 357-484 (Academic Press, London, 1974).

8. Wilson, K.G. Renormalization group and critical phenomena. I. Renormalization group and the Kadanoff scaling picture. *Phys.Rev.B* **4**, 3174-3205 (1971).

9. Kadanoff, L.P. *et al.* Static Phenomena Near Critical Points: Theory and Experiment. *Rev.Mod.Phys.* **39,** 395-437 (1967).

10. Hill, T. L. *Statistical Mechanics – Principles and Selected Applications* (McGraw-Hill Book Company, USA, 1956).

11. McCoy, B.M. & Wu, T.T. *Two-Dimensional Ising Model* (Harvard University Press, Cambridge, Mass, 1973).

12. Pushpalatha,K. & Sangaranarayanan.M.V. Two-dimensional condensation of organic adsorbates at the mercury/aqueous solution interface: A global analysis of critical parameters using Ising model formalism. *J.Electroanal.Chem*.**425,** 39-48 (1997).

13. Saradha, R. & Sangaranarayanan, M. V. Theory of electrified interfaces : A combined analysis using density functional approach and Bethe Approximation. *J.Phys.Chem B.* **102,** 5468-5474 (1998).

14. Aoki, K. Theory of phase separation of binary self-assembled films. *J.Electroanal.Chem.* **513,** 1-7 (2001).

15. Lenz, P., Zagrovic, B., Shapiro, J. & Pande, V.S. Folding probabilities: A novel approach to folding transitions and the two-dimensional Ising-model. *J.Chem.Phys.* **120,** 6769-6778 (2004).

16. Kole, P. R., de Vries, R. J., Poelsema, B. & Zandvliet, H. J. W. Free Energies of steps on (1 1 1) fcc surfaces. *Solid State Comm.* **136,** 356-359 (2005).

17. Thompson, C. J., *The Annals of Probability*. **14**, 1129-1138 (1986).